\documentclass[pre,preprint,floatfix,amsmath,amssymb]{revtex4}
\usepackage{graphicx}
\usepackage{dcolumn}
\include{epsf}

\newcommand{\be}{\begin{equation}}
\newcommand{\ee}{\end{equation}}

\newcommand{\vomega}{\mbox{\boldmath $ \omega $}}
\newcommand{\vOmega}{\mbox{\boldmath $ \Omega $}}
\newcommand{\vmu}{\mbox{\boldmath $ \mu $}}

\begin{document}
\title{Magnetic field reversals and long-time memory in conducting flows}

\author{P. Dmitruk$^1$, P.D. Mininni$^1$, A. Pouquet$^2$, S. Servidio$^{3}$ and W.H. Matthaeus$^4$}
\affiliation{$^1$ Departamento de F\'isica, Facultad de Ciencias Exactas y Naturales, Universidad de Buenos Aires and IFIBA, CONICET, Buenos Aires, Argentina\\
$^2$ Department of Atmospheric and Space Physics, University of Colorado 
and National Center for Atmospheric Research, Boulder, Colorado, USA\\ 
$^3$ Dipartimento di Fisica, Universita della Calabria, Cosenza, Italy\\
$^4$ Bartol Research Institute and Department of Physics and Astronomy, University of Delaware, Newark, Delaware, USA}

\begin{abstract}
Employing a simple ideal magnetohydrodynamic model in spherical 
geometry,
we show that the presence of 
either rotation or finite magnetic helicity is 
sufficient to induce dynamical reversals of the magnetic dipole
moment. The statistical character of the model is similar to that 
of terrestrial 
magnetic field reversals, with the similarity being stronger 
when rotation is present. 
The connection between long time correlations, $1/f$ noise,
and statistics of reversals is supported,
consistent with earlier suggestions. 
\end{abstract}
\pacs{47.65.-d 47.27.E- 47.35.Tv 91.25.Mf}

\pacs{}
\maketitle

\section{Introduction}
The origin of magnetic field reversals in the Earth magnetic field is 
a matter of debate. Reversals take place rapidly, within
a scale of $\sim 1000$ years, but infrequently, 
distanced apart by periods of $10^4$--$10^7$ years
\cite{CandeKent95,ValetEA05}.

Reversals, first thought to be a purely random process, are now known 
to display long-term memory with deviations from a purely Poisson 
process 
\cite{CarboneEA06}.
In many systems, 
such a 
high degree of variability may be associated with features 
of the power spectrum of the time series known as
``$1/f$'' noise, 
an indication of the presence 
of correlations over a 
wide range of time scales
\cite{vanderZiel50,Machlup,MontrollShlesinger,DmitrukMatthaeus07}.
$1/f$ 
signals are found in 
many
physical systems 
\cite{DuttaHorn81,WestShlesinger89},
including the 
intensity of the 
geomagnetic field \cite{ZieglerConstable11,ConstableJohnson05}.
We examine
this phenomenon 
by employing a simple model consisting 
of incompressible 
non dissipative magnetohydrodynamics (MHD) 
in spherical geometry.
The
results demonstrate the presence of
reversals possessing $1/f$ noise 
where rotation and/or magnetic helicity 
play important roles. 
Remarkably, the distribution of waiting
times between reversals follows a power law 
that is comparable to the record of 
terrestrial magnetic reversals. 

The Earth's magnetic field is sustained by a dynamo process:
motions of the conducting fluid core
generate and sustain magnetic 
fields against Ohmic dissipation. 
Although MHD contains 
the basic physics of the dynamo, 
the complete terrestrial problem 
requires solving 
either compressible, Boussinesq, or anelastic MHD equations for the
velocity, the magnetic field, 
and the temperature, in a spherical shell with a 
possible inner solid conducting core, and surrounded by a mantle 
\cite{GlatzmaierRoberts95,AmitEA10}. Additional realism 
requires more complexity in
chemistry, equations of state, and boundary conditions.
Even with advanced supercomputers, only few reversals can be simulated
\cite{GlatzmaierRoberts95,AmitEA10,OlsonEA11}, and studies of the 
long-time statistics of reversals are therefore out of reach.

Experiments reproducing
dynamos in laboratory turbulent flows 
display magnetic 
field reversals \cite{BerhanuEA07,BaylissEA07}, 
$1/f$ noise and long-term memory.
Still, 
a theoretical understanding of these features remains incomplete
since the origin of 
 correlations
with time scales much greater than 
the characteristic nonlinear time associated with the 
largest eddies in the system is unknown.  

Many physical causes 
have 
been considered to explain the origin and statistics of the reversals,
including the effect of tides, departures of the mantle from 
spherical geometry, or low magnetic Reynolds number effects. 
We 
show that the MHD equations in their simplest nonlinear form 
(incompressible and ideal) in a simple geometry 
(spherical surrounded by a perfect conductor) already include the
ingredients required for 
magnetic field reversals, long-time
correlations, $1/f$ noise, and non-Poisson statistics compatible with
that observed in the geodynamo.

\section{Model}

The ideal MHD
equations 
are solved using a spectral method 
that preserves to numerical accuracy all 
ideal quadratic invariants 
with no numerical dissipation or dispersion. 
For very long time integrations, 
this is the only method that ensures adequate conservation. 
Since the initial energy introduced in the system is conserved, no 
external forces are needed to sustain the velocity and magnetic fields. 
For a purely spectral Galerkin method, 
we use 
spherical Chandrasekhar-Kendall functions as a basis, 
expanding the fields in 
spectral space 
 \cite{Mininni06,Mininni07}). A 
fourth-order Runge-Kutta method is used to evolve 
the system in time.

With the magnetic 
field confined in the interior of the sphere, the system has two 
quadratic conserved quantities: the total energy (kinetic plus 
magnetic, $E=\frac{1}{2}\int |{\bf v}|^2 + |{\bf b}|^2 \, dV$,
with ${\bf v}$, ${\bf b}$ the velocity and magnetic fields) 
and the magnetic helicity ($H_m=\int {\bf a} \cdot {\bf b} \, dV$, a measure of 
linkage or handedness of the magnetic field,
with ${\bf a}$ the vector potential, $\nabla \times {\bf a} = {\bf b}$). 
$E$
is transferred towards small scales (``direct cascade''), 
while $H_m$ is transferred towards large scales (``inverse cascade'').
In the ideal system, 
$H_m$ condenses at the largest 
available scales \cite{FrischEA75}. 
In our simulations, 
long timescale correlations  
arise when 
$H_m$ is non-zero.
Long time 
correlations 
also arise due to symmetry breaking
by rotation \cite{MininniEA11}. 
Here we show
for the rotating sphere, 
that the magnetic dipole moment 
reverses with respect to the rotation direction,
displaying $1/f$ noise and long-term memory even 
when the magnetic helicity is zero.
A recent related study \cite{Shebalin12} 
reported 
persistence of the magnetic dipole 
associated with broken ergodicity effects \cite{Shebalin89}.
Broken ergodicity of fluid systems 
may also be viewed as ``delayed ergodicity'' in which 
very long times correlations delay 
ergodically 
covering the phase space
\cite{ServidioEA08}.

The incompressible ideal MHD equations solved
for the evolution of the velocity field ${\bf v}$
and magnetic field ${\bf b}$ (in Alfvenic units) are
\begin{equation}
\frac{\partial {\bf v}}{\partial t} = {\bf v} \times \vomega + 
    {\bf j} \times {\bf b} - \nabla \left({\mathcal P} + \frac{v^2}{2} 
    \right) - 2 \vOmega \times {\bf v} , 
\label{eq:momentum}
\end{equation}
\begin{equation}
\frac{\partial {\bf b}}{\partial t} = \nabla \times \left( {\bf v} 
    \times {\bf b} \right) ,
\label{eq:induction}
\end{equation}
for 
vorticity $ \vomega = \nabla \times {\bf v}$;
electric current density 
${\bf j} = \nabla \times {\bf b}$;
normalized pressure 
${\mathcal P}$;
and rotation rate $\vOmega$. 
The units are normalized 
to the spherical radius $R$ and 
initial root mean
square velocity $v_0=\left<{\bf v}^2\right>^{1/2}$,
so $R=1$ and $v_0=1$, and 
the time unit is 
$t_0=R/v_0=1$ (later 
timescales are rescaled to Ma $=10^6~{\rm years}$,
based on the longest observed 
waiting time between reversals).
We consider vanishing normal velocity and magnetic field components
at the sphere boundary. 
For the simulations, $980$ coupled 
Chandrasekhar-Kendall (C-K) modes are 
followed in time.
The C-K functions are
\begin{equation}
{\bf J}_i = \lambda \nabla \times {\bf r} \psi_i + \nabla \times \left(
    \nabla \times {\bf r} \psi_i \right) ,
\label{eq:ji}
\end{equation}
where we work with a set of spherical orthonormal unit vectors
$(\hat{r},\hat{\theta},\hat{\phi})$, and the scalar function $\psi_i$ is a
solution of the Helmholtz equation, $(\nabla^2 + \lambda^2) \psi_i = 0$.
The explicit form of $\psi_i$ is
\begin{equation}
\psi_i (r, \theta, \phi) = C_{ql} \, j_l(|\lambda_{ql}| r) Y_{lm}
    (\theta,\phi),
\label{eq:psi}
\end{equation}
where $j_l(|\lambda_{ql}| r)$ is the order-$l$ spherical Bessel function of the first
kind, $\{ \lambda_{ql}\}$ are the roots of $j_l$ indexed by $q$
(so that the function vanishes at $r=1$),
and $Y_{lm}(\theta,\phi)$ is a spherical
harmonic in the polar angle $\theta$ and the azimuthal angle
$\phi$.
The sub-index $i$ is a shorthand notation for the three indices $(q,l,m)$;
$q=1,2,3,\dots$ corresponds to the positive values of $\lambda$, and
$q=-1,-2,-3,\dots$ indexes the negative values; finally $l=1,2,3,\dots$,
and $-l \le m \le l$. The C-K functions satisfy
\begin{equation}
\nabla \times {\bf J}_i = \lambda_i {\bf J}_i  \ .
\end{equation}
With the proper normalization constants, they are a complete orthonormal set.
The values of
$|\lambda_i|$ play a role similar to the wavenumber $k$ in a Fourier
expansion. Note that the boundary conditions, as well as the Galerkin method to
solve the equations inside the sphere using this base, were chosen to
ensure conservation of all quadratic invariants of the system (total
energy and magnetic helicity).

The initially excited modes for the runs are those for
$q=\pm 3$, $l=3$ and all possible values of $m$.
With proper initial values for the expansion coefficients of
the C-K functions, the
initial values of the quadratic quantities can be chosen.
In all the runs, the initial total energy is set to $E=1$ (dimensionless units).
We set the initial magnetic and kinetic energies to $E_m = E_k \approx 0.5$.
The runs with non-zero magnetic helicity have $H_m \approx 0.03$.
As a comparison, note that for the $q=3$, $l=3$ mode alone, $H_m/E_m$
is no more than about $0.072$ (this is the maximum value of $|H_m/E_m|$
if only modes with $|q|=3$, $l=3$, and one sign of $\lambda$
are excited). So, the chosen value of $H_m$ (when is non-zero)
corresponds to about $85\%$ of the maximum helicity in the
system.

\section{Results}
\subsection{Kinetic and magnetic energy, field structure}
The values of the total energy $E$ and magnetic helicity remain constant
in time (as ideal invariants). The values of the magnetic and
kinetic energies $E_m$ and $E_k$ fluctuate, reaching a statistical
steady state after about $20$ unit times. The initial ratio of
kinetic over magnetic energy is $E_k/E_m(t=0)=1$ and
approachs and fluctuates around $E_k/E_m \approx 0.9$.
We performed two additional runs starting from different initial ratios
$E_k/E_m (0)\approx 2$ and $E_k/E_m (0) \approx 0.5$, with the
same value of magnetic helicity $H_m=0.03$. Both cases evolve initially
and after about $20$ unit times reach the same asymptotical statistical state,
with a value of $E_k/E_m \approx 0.9$. This is shown in Fig. 1.
This asymptotic value corresponds to a steady
state with some excess of magnetic energy over kinetic energy which
is consistent with the non-zero value of magnetic helicity (which allows
condensation at the large scales).

\begin{figure}
\includegraphics[width=\linewidth]{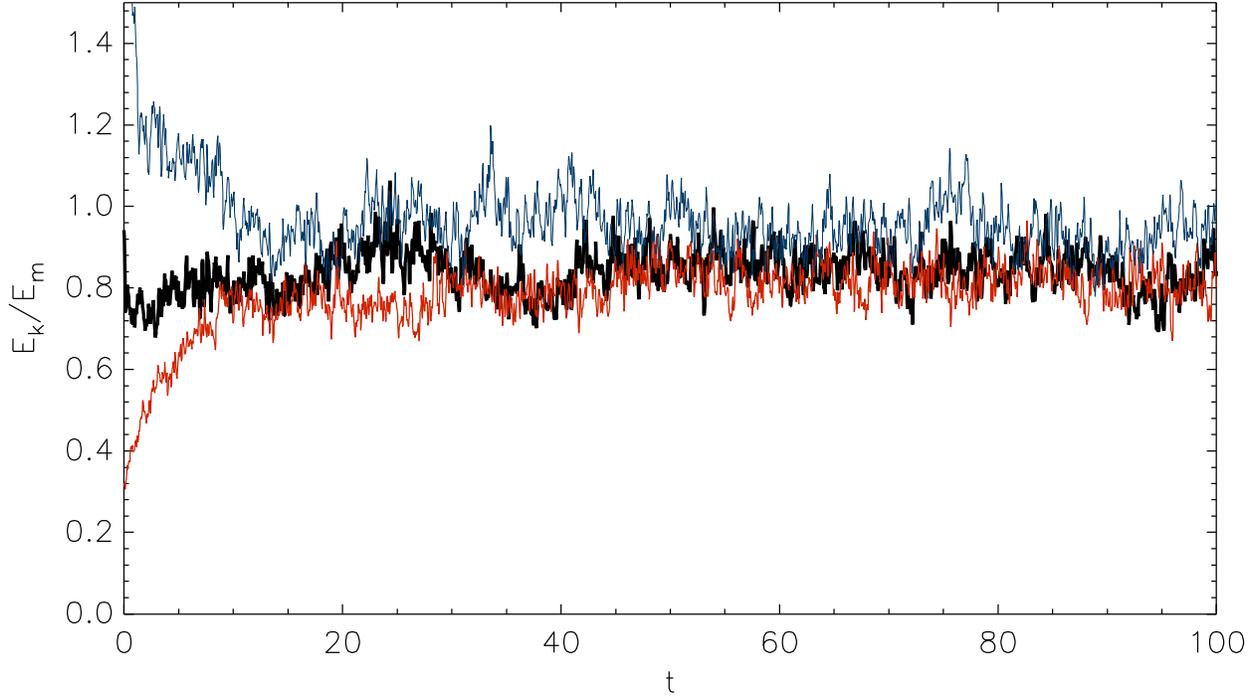}
\caption
{
The ratio of kinetic energy vs magnetic energy $E_k/E_m$ as a function of
time, for three runs with different initial conditions and same
$H_m=0.03$, $\Omega=16$. Thicker line $E_k/E_m(t=0)=1$, 
intermediate thick line $E_k/E_m(t=0)=0.5$, thin line
$E_k/E_m(t=0)=2$.
}
\label{fig:ratio_ekem}
\end{figure}

The results about the statistics of the magnetic dipole that follows (next 
subsection) are not sensitive to the different initial values of 
the ratio $E_k/E_m$.

The fields evolve to a highly disordered state,
with a wide range of scales present. Fig. 2 shows velocity and
magnetic field lines for one particular run.

\begin{figure}
\includegraphics[width=5cm]{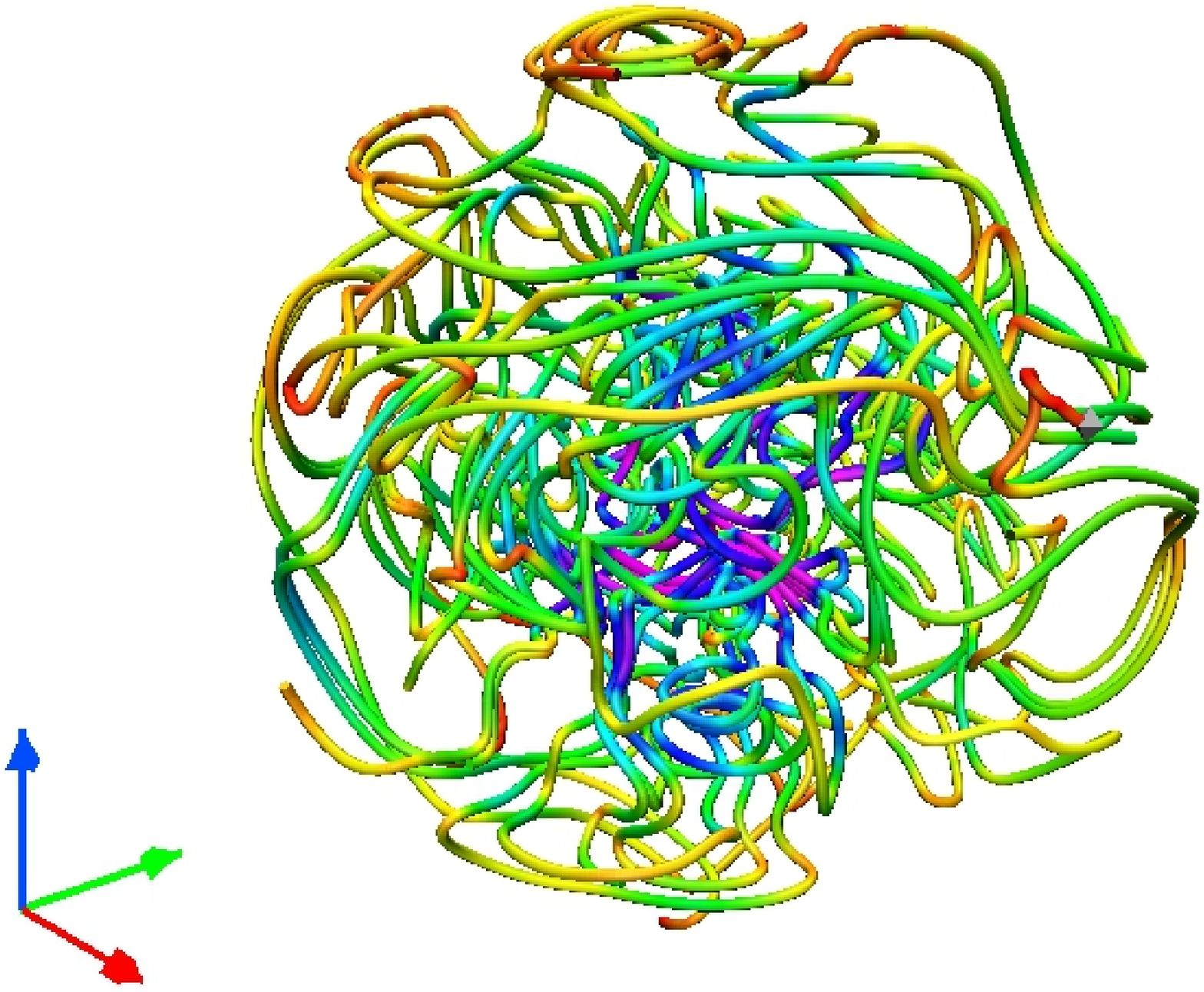}
\includegraphics[width=6.4cm]{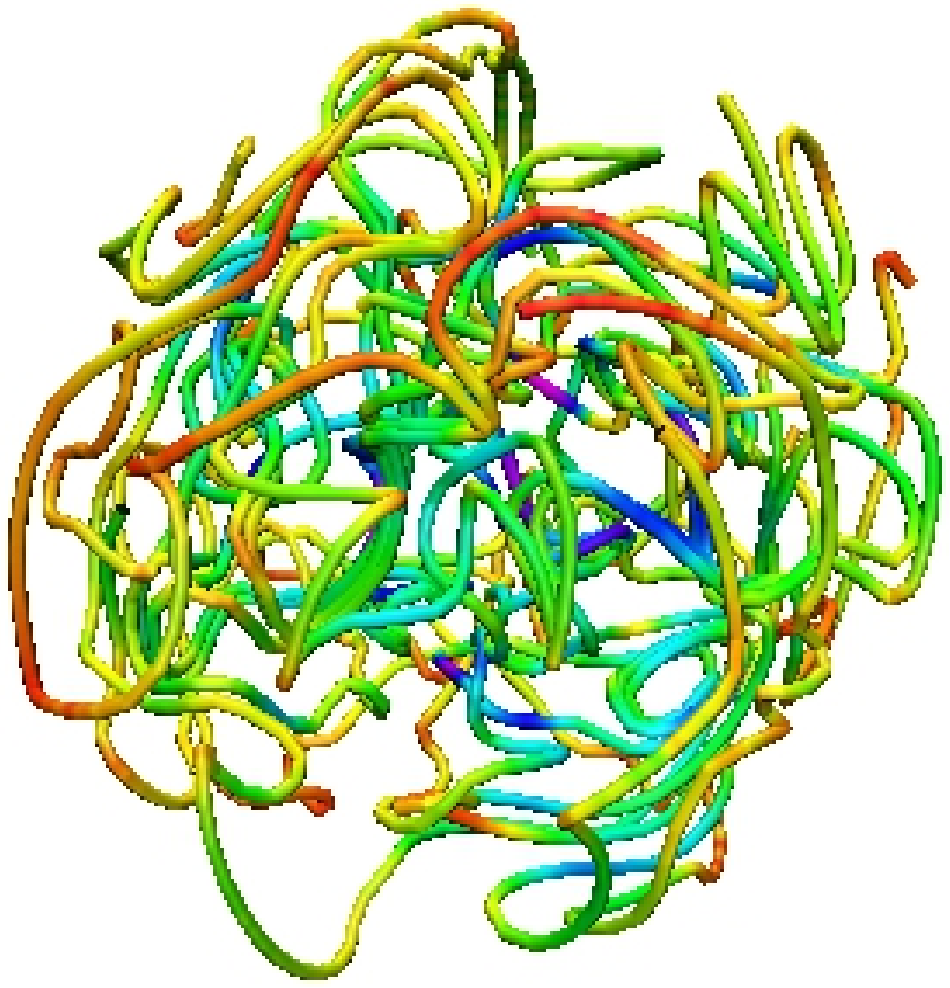}
\caption
{
Velocity (top) and magnetic (bottom) field lines in the run
with $\Omega=16$, $H_m=0.03$. The field lines
change color according to the intensity of the field, from red to
yellow, blue and magenta. The red, green and blue arrows indicate respectively
the $x,y,z$ axis, with $\Omega$ in the $z$-direction.
}
\label{fig:fields}
\end{figure}

\subsection{Magnetic dipole and statistics of reversals}
We focus on the dynamics of the magnetic dipole moment
\begin{equation}
\vmu = \frac{1}{2} \int {\bf r} \times {\bf j} \, dV \, ,
\end{equation}
and in particular, on its
$z$-component $\mu_z$,
which is of importance 
with rotation $\vOmega = \Omega \hat{z}$.
We report first results of a run with both 
non-zero magnetic helicity ($H_m=0.03$)
and non-zero rotation ($\Omega=16$ in units of $t_0^{-1}$
defined above).
Figure 3 (top panel) 
shows the time evolution of the z-component of 
the dipole moment $\mu_z$. 
The simulation extends for $5000 t_0$ 
but for clarity only the
segment from $t=50$ to $t=300$
is illustrated.
The sign of $\mu_z$
changes many
times during this period; 
many reversals are observed. 
The time periods between reversals range from short
times
($\delta t \sim 1$) to long times ($\delta t \sim 50$). 

\begin{figure}
\epsfxsize=8cm
\centerline{\epsffile{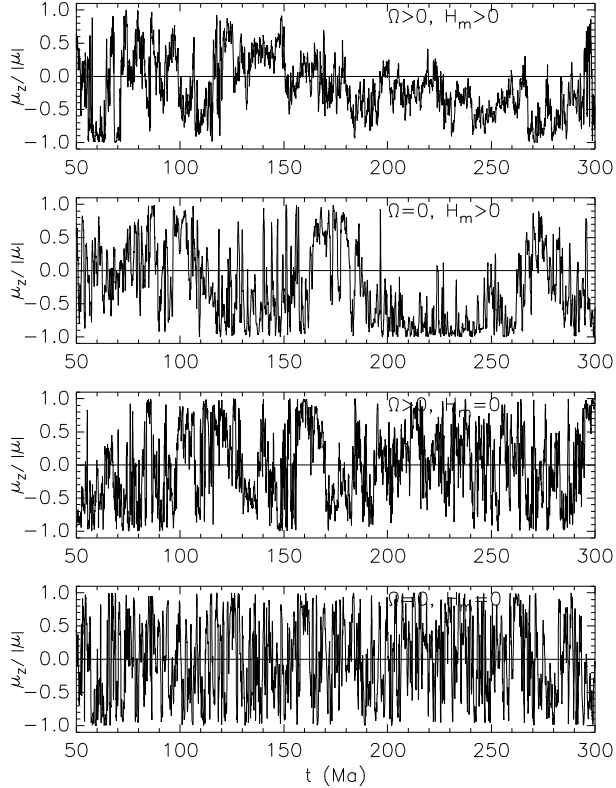}}
\caption{
Time series of the normalized magnetic dipole moment for four different 
values of the angular velocity of rotation $\vOmega$ and magnetic 
helicity $H_m$. Time is measured in units of Ma (mega anni), 1 Ma = $10^6$
years, as indicated in the text.
}
\end{figure}

The statistics
of these fluctuations are 
analyzed by computing the frequency
power spectrum $P(f)$, shown in Figure 4 (top). 
The spectrum 
is obtained by Fourier transforming the $\mu_z(t)$
time series in 10 non-overlapping samples, 
averaging the estimates of 
$P(f)$ to improve statistics. 
The frequency 
$f=0.5$ 
corresponds to the longest nonlinear time scale 
that can be constructed based on local 
dimensional arguments, using 
the longest available scale in the system $2R$,  and
a unit root mean square velocity. 
For 
a system with no long term 
memory effects, $P(f)$ would 
be flat (constant) at lower frequencies, 
corresponding to uncorrelated fluctuations 
at time scales 
longer than the autocorrelation time $t_c=2R/v_0=2$. 
However, 
the substantial 
excess power at frequencies $f < 1/t_c = 0.5$ 
indicates a long term memory 
not controlled by a single correlation time. 
This effect is known as $1/f$ noise, 
corresponding to  the typical (approximate) 
power law spectrum found at the low 
frequencies \cite{Machlup,WestShlesinger89}.
A $1/f$ power law is illustrated 
in Figure 4.
The appearance of $1/f$ noise in
ideal 
fluid models
has been discussed in
\cite{DmitrukEA11}.
The inset included in Fig. 4 corresponds to the compensated spectrum, 
that is $f P(f)$, which should be flat for a $1/f$ spectrum. 
This plot indicates clearly the wide range of frequencies for which
we can see a $1/f$ in this case.

\begin{figure}
\epsfxsize=5.5cm
\centerline{\epsffile{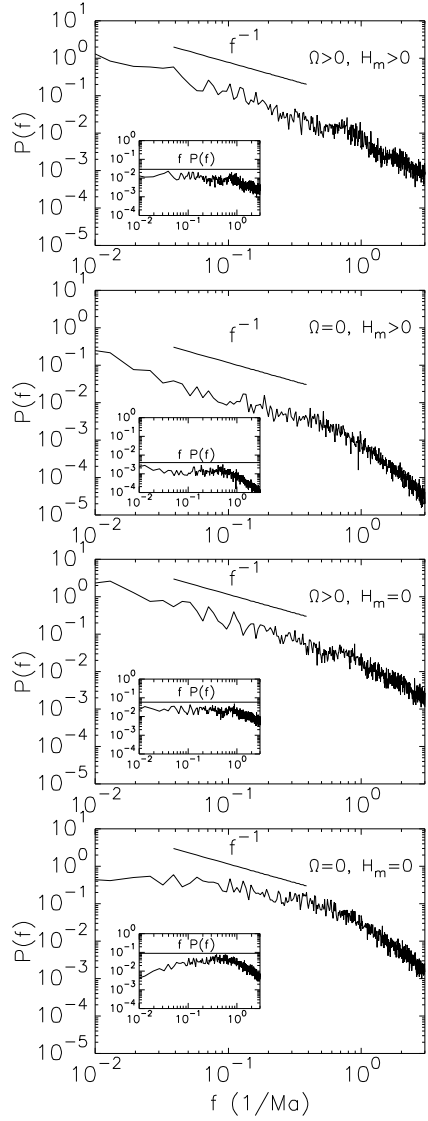}}
\caption{
Frequency spectra of the magnetic dipole moment for four different values of the angular velocity of rotation $\vOmega$ and magnetic helicity $H_m$. 
A $f^{-1}$ power law spectrum is indicated as a reference in each plot.
Also, insets show the compensated spectra $f P(f)$ for each case.
Units of frequency are 1/Ma, 1 Ma = $10^6$ years.
} 
\end{figure}

To quantify reversals,
we compute statistical distribution
of times between reversals 
of $\mu_z$, 
i.e., 
the {\it waiting time
distribution}
\cite{CarboneEA06}. 
The probability distribution function (PDF) of the
waiting times 
obtained from the simulation
is shown in Figure 5 (top).
Also shown in Figure 5 is the 
known distribution of waiting times from data measurements
of the geomagnetic reversals in \cite{CandeKent95}.
To compare these,
we arbitrarily 
identify the longest simulation 
waiting time
with the longest reported waiting time
for geomagnetic reversals. The latter is 
$\sim$ 30 Ma. 
The relevant point 
here is that the same trend is observed 
for the waiting times -- this corresponds to a power law, 
indicating 
the existence of long term memory and non-poissonian statistics 
\cite{CarboneEA06,SorrisoEA07}. 
This long term memory
is associated 
with 
the $1/f$ noise observed in the
power frequency spectrum (Figure 4).

\begin{figure}
\epsfxsize=6cm
\centerline{\epsffile{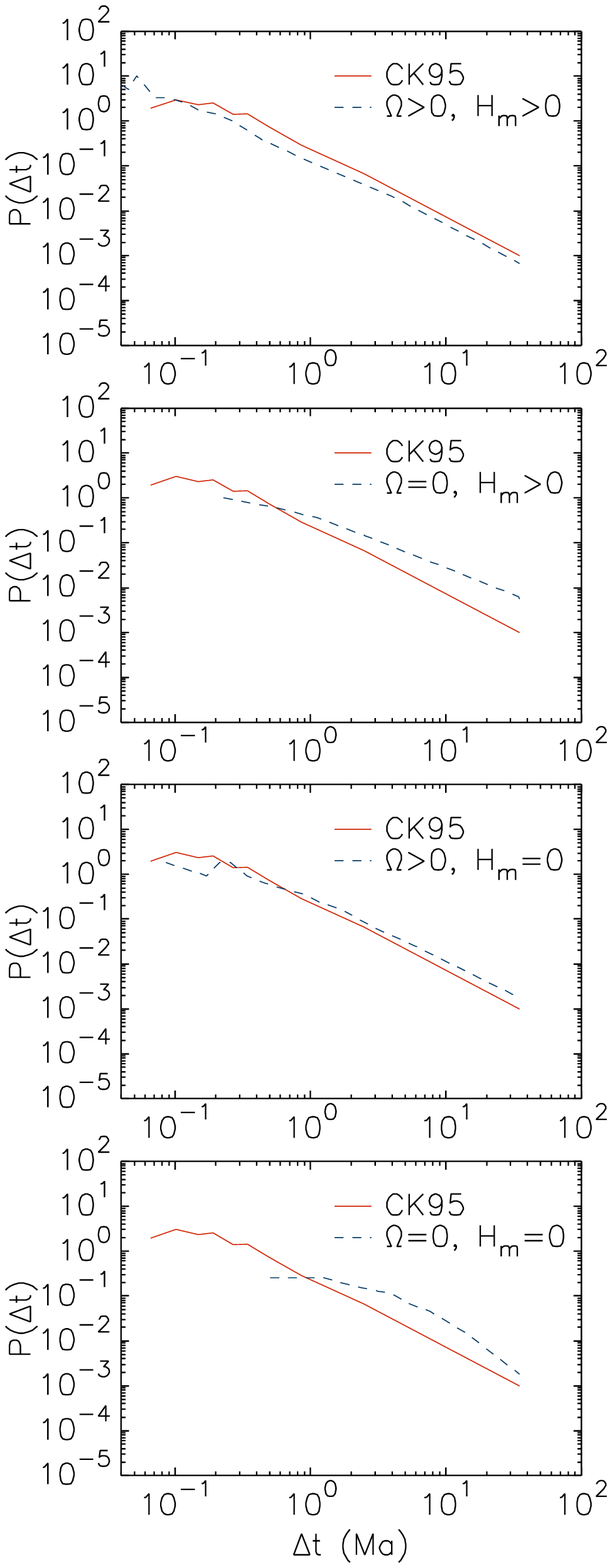}}
\caption{
Distribution function of the waiting time between reversals, 
for four different values of the angular velocity of rotation $\vOmega$ 
and magnetic helicity $H_m$ (dashed lines). The continuous line 
corresponds to 
the distribution function for the observational Cande and Kent 1995 data
\cite{CandeKent95}
}
\end{figure}

Next, we show results with non-zero magnetic helicity $H_m=0.03$
but no rotation ($\Omega=0$).
The 
dipole moment time
series, frequency spectrum and waiting time distribution 
are shown in the second panels of Figures 3, 4 and 5 respectively. 
Although the frequency spectrum 
shows that there is still an excess power at 
$f < 1/t_c$,
this effect is weaker than 
in the case with both rotation and magnetic helicity.
The compensated spectrum in the inset of this Figure also indicates
the range of frequencies for which a $1/f$ is observed.
In addition, the distribution of waiting times 
departs more from 
the observational data of Cande and Kent 1995 \cite{CandeKent95}.

Results with zero magnetic helicity ($H_m=10^{-7}$) and  
maintaining rotation ($\Omega=16$) are shown in the third panels of 
Figs. 3, 4 and 5 respectively.
These results are similar to the first case of non-zero rotation and
magnetic helicity, showing 
excess power at low frequencies (Fig. 4), flat compensated spectrum
(inset)
and comparable results with the observational data for the
waiting time distribution (Fig. 5).

Finally,
we present results with zero magnetic helicity ($H_m=10^{-7}$) and  
no rotation ($\Omega=0$).
(Bottom panels of Figures 3, 4 and 5 respectively).
In this case, the absence
of excess power at lower frequencies is clear.
This is more visible in the compensated spectrum plot (inset). 
Also, a larger departure from the observational data 
(Fig 5)is noted.

\subsection{The Hurst exponent}
In order to have another measure of comparison
we have additionally computed the
Hurst exponent $H$
for each of the time series. 

The standard definition is that 
a process $g(t)$ is self-similar, with self-similarity (Hurst) exponent $H\in(0,1)$, 
if it satisfies 
\begin{equation}
g(\lambda t)\sim \lambda^H g(t)   ~~~(\lambda>0).
\label{eq:self}
\end{equation}
Even if a process is nonstationary, it can satisfy Eq. \ref{eq:self})
if the increment process $\delta g=g(t)-g(t+1)$ is stationary.
A canonical example
is Brownian motion, for  which $H=1/2$.
When $1>H>0.5$, there are
long-range time correlations (persistence), when $0.5>H>0.0$, 
the series has long-range anticorrelations (antipersistence), and, 
for a time series with no long time correlations, $H=0.5$.
The $H$ parameter \cite{hurst51,Mandelbrot69}
is often used to characterize
long-range dependences. 

\begin{figure} 
\epsfxsize=8cm
\centerline{\epsffile{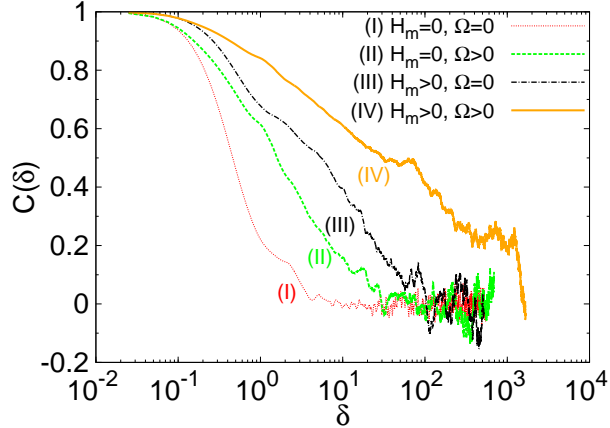}}
\caption{Autocorrelations functions vs. 
time-lag $\delta$.}
\label{corrcomp}
\end{figure}

Long range memory is also 
connected with powerlaw 
decay of the autocorrelation function, 
with index $\beta$, 
\begin{equation}
    C(\delta) = \langle \mu_z(t)\mu_z(t+\delta) \rangle \sim \delta^{-\beta}.
\label{eq:c}
\end{equation}
here written for the magnetic moment $\mu$.
The autocorrelation function, for all the 
cases in our paper are compared in Fig. \ref{corrcomp}.
This indicates clearly the 
relevance of magnetic helicity and rotation. 
In fact, where both are present,
the tail of the correlation function seems to be more power-law like.

For cases with long memory, 
computation of the 
power spectral density (PSD) becomes difficult; 
however when the PSD displays a low frequency 
powerlaw range,  the associated Hurst exponent is
also found as
\begin{eqnarray}
\nonumber
P(f)\sim f^{-\alpha}, \\
1<\alpha=2H+1<3.
\label{eq:sp}
\end{eqnarray}
In principle, the above can be used 
to  estimate $H$. 
Performing a fit of the power spectra at low frequencies, 
we obtained the spectral indexes $\alpha$
for each case, reported in Table I. 
A different method, described below, 
is used to obtain $H$.

\begin{table}
\caption{Hurst analysis of the dipole moment
for simulations with different 
$H_m$ and $\Omega$.
Fit to the power spectrum is $\alpha$ 
(as in Eq. \ref{eq:sp}). 
Hurst exponent $H$ is an
average over 
$H_q$ in Eq. (\ref{eq:sqa}) .
}
    \label{tab:Vlasov}
\begin{center}
\begin{tabular}{c c | c | c }
\hline
  $H_m$ & $\Omega$ & $\alpha$ &   $H$ \\
\hline
 $>0$ & $>0$ & $1.13\pm0.06$  &   $0.107\pm 0.007$ \\
 $>0$ & $=0$ & $1.1\pm0.1$    &   $0.127\pm 0.008$ \\
 $=0$ & $>0$ & $0.8\pm0.1$    &   $0.05\pm 0.01$  \\
 $=0$ & $=0$ & $0.3\pm0.1$    &   $0.010\pm0.004$ \\
\hline
\end{tabular}
\end{center}
\end{table}

\begin{figure} 
\epsfxsize=7cm
\centerline{\epsffile{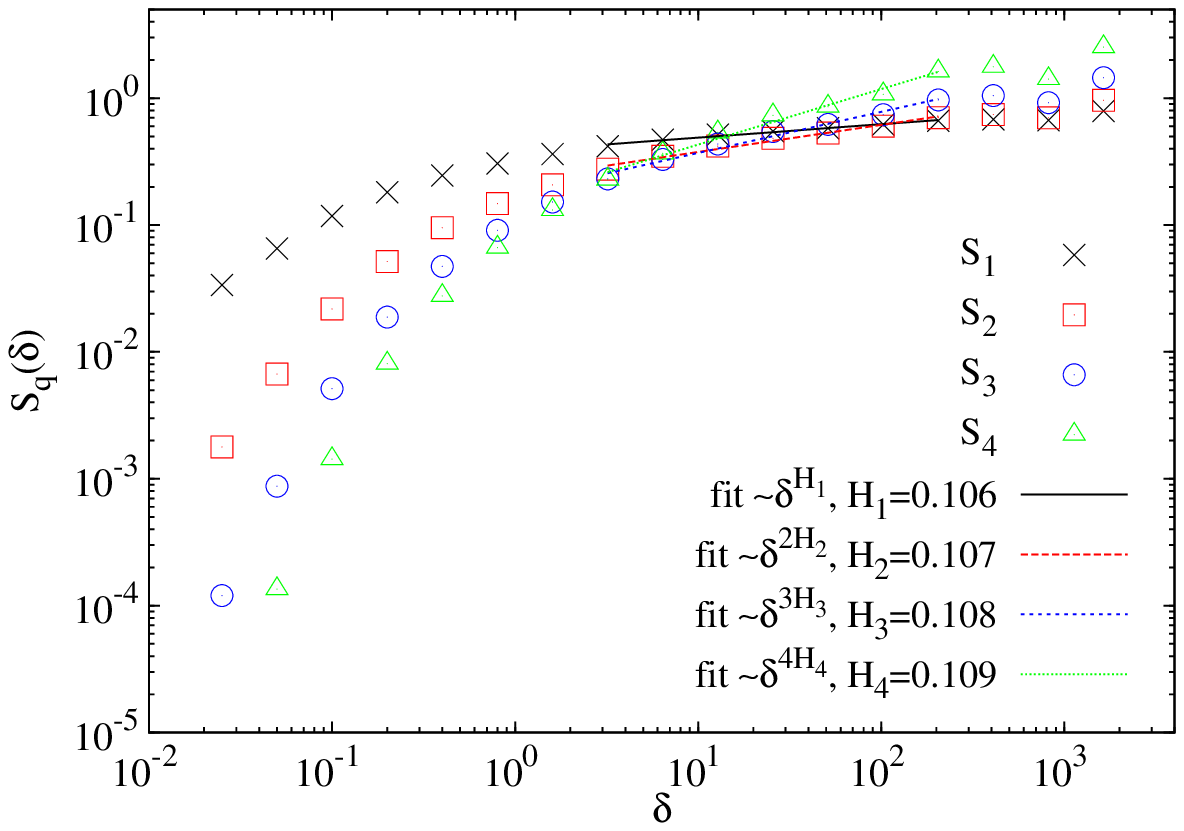}}
\caption{Hurst analysis for $H_m>0$, $\Omega>0$. The structure functions, computed up to the $4^{th}$ moment, 
are represented with open symbols, while the fits, 
from Eq. (\ref{eq:sqa}), are reported with lines. In the 
legend of the plot, the results of the fit $H_q$ are reported as well.}
\label{allHW}
\end{figure}

The main technique we
use is based on structure function 
analysis in which 
the expected size of changes for time lag $\delta$ 
are related to the exponent $H$.
For the $q^{th}$ order structure function
one obtains an independent  estimate $H_q$, 
where
\begin{equation}
S_q(\delta) = \langle \left| \mu_z(t+\delta)-\mu_z(t) \right|^q \rangle \sim \delta^{\zeta(q)}\equiv \delta^{q H_q}.
\label{eq:sqa}
\end{equation}
A general statement \cite{jgr}, 
can be made about the family of exponents: $\zeta(q)$ will be concave, 
$d^2 \zeta/d ^2<0$. If the signal has absolute
bounds, it can be shown that $\zeta(q)$ is monotonically 
nondecreasing \cite{Frisch}. 
Concavity
alone is sufficient to define a hierarchy of exponents $\zeta(q)=q H_q$.

The relation between definitions in Eqs. (\ref{eq:self}), (\ref{eq:sp})
and (\ref{eq:sqa}) is not immediate, indeed it has some subtleties,
clearly documented in the literature. For a good description of
the problem, see for example \cite{Gilmore,Flandrin}. A typical example
of self-similar process is given by the fractional brownian motion (FBM),
which can be regarded as a generalization of the well-known brownian motion
which has $H=1/2$. Although the power spectral density is not defined for
a non-stationary self-similar process such as FBM, it has been shown that
a time-averaged power spectra satisfy the relation in Eq. (\ref{eq:sp}),
by means of a time-frequency analysis. An explanation can be found
in \cite{Flandrin}. 

Fig. \ref{allHW} 
shows a Hurst analysis for the run with $H_m>0$ and $\Omega>0$. 
Structure functions, computed for $q=1,2,3,4$, 
reveal that at large $\delta$
(low frequencies), a self-similar 
scaling is present. 
As stated in Eq. (\ref{eq:self}), this behavior is typical of monofractal signals.
Note that the range of scales chosen for the fits are comparable 
to the range of the $1/f$ noise 
in the power spectrum. 
The higher order structure functions give results 
consistent with monofractality in that $H_q$ 
is independent of $q$ (see Eq. \ref{eq:sqa}).
The break point of the large scale noise 
is roughly at $\delta\sim 1$, the nonlinear time.
Finally, we estimate 
$H = \frac{1}{4}\sum_{q=1}^4 H_q$, 
and the results are shown 
in Table I.

\begin{figure} 
\epsfxsize=7.5cm
\centerline{\epsffile{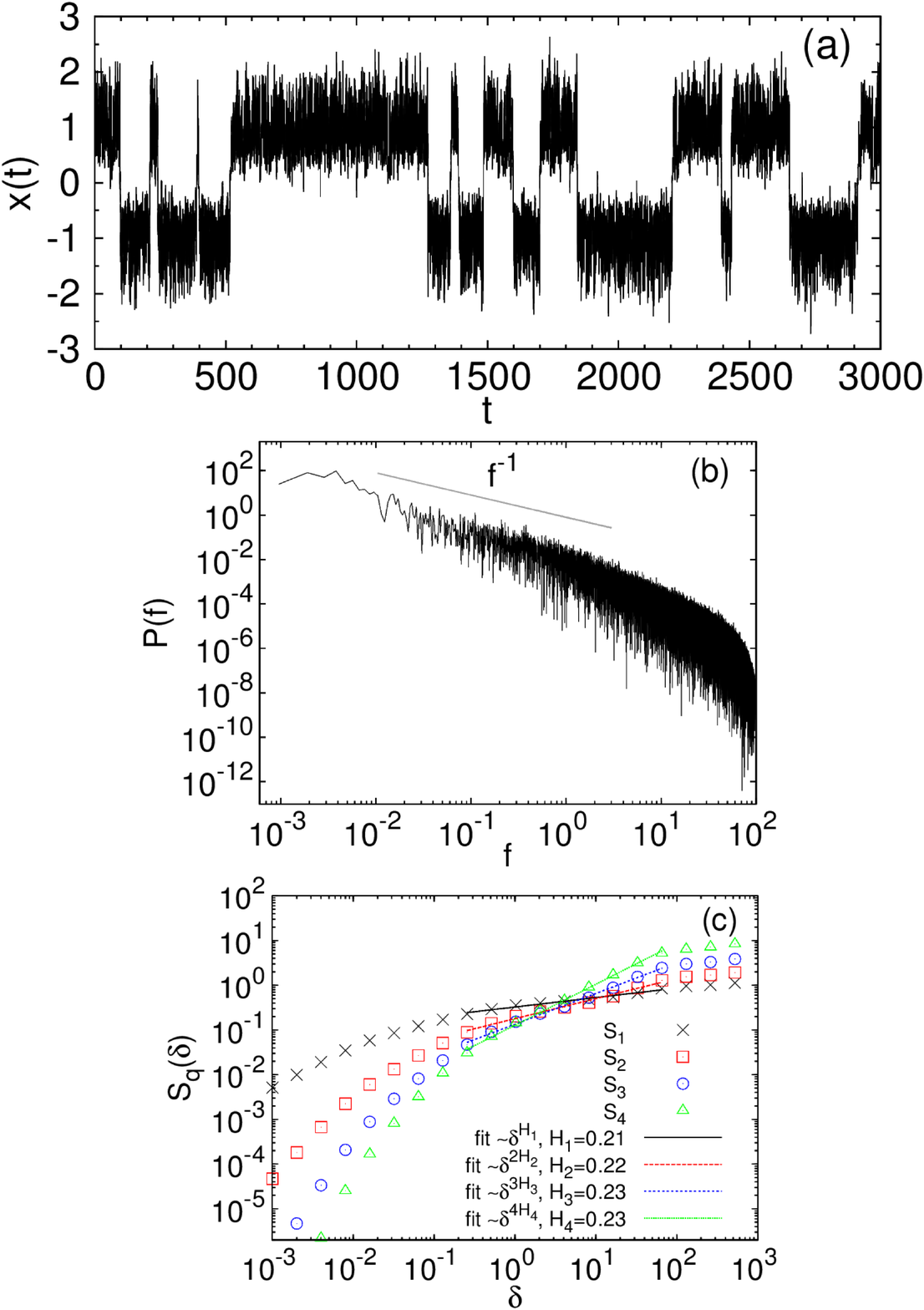}}
\caption{Hurst analysis for HD model: (top) time series, (middle) power spectrum;
(bottom) structure function analysis.}
\label{hdfig}
\end{figure}

The case $H_m=0$, $\Omega=0$ is very particular, 
having a flat spectrum, while its Hurst exponent is 
small and consistent with zero, 
typical of white noise.
On the contrary, 
both the cases with ($H_m>0,\Omega>0$), and ($H_m>0,\Omega=0$) have $H\sim 0.1$, indicating that the signal is anti-persistent.

At this point we may 
ask if our Hurst analysis results  
are comparable to that of geomagnetic reversals. Unfortunately, 
this analysis cannot be performed on simplified data such as the CK95 dataset.
To have an independent estimate of $H$ for these systems, we make use of a 
simple model of geomagnetic dynamo, proposed by 
Hoyng and Duistermaat (HD) \cite{refhd}. Very briefly, 
the HD model, inspired by bistable chaotic systems, describes 
the axysimmetric component of the dynamo field.
The nonlinear evolution takes into account the
back-reaction of the Lorentz force on the flow. 
After simplifications the model reduces 
to a multidimensional bistable
oscillator driven by multiplicative noise 
[see Eq.s (2)-(4) of \cite{refhd}].
We solved those model equations numerically, and obtained 
the time series using the same parameters as in \cite{refhd}. 
As reported in Fig. \ref{hdfig} (top), 
the solution manifests strong similarities with the geomagnetic
reversals.  
The power spectrum (middle panel)
exhibits a power-law consistent with $1/f$ noise (slightly steeper.)
Finally, generalized Hurst analysis (bottom)
is reported, showing that for this simple geomagnetic model,
$H\sim0.2$. 
This result is close to the Hurst exponent 
of dipole moment derived from our simulations,
and suggests that geomagnetic reversals are likely to have $0<H<0.5$,
typical of systems with long-range anticorrelations.

\section{Conclusions}
We obtained reversals of the dipole magnetic field with a
simple ideal MHD system, in spherical geometry
and no externally imposed driving. 
Numerical simulations for durations of 
5000 nominal times were performed, 
for cases including or not including rotation, and
for cases including or not including magnetic helicity.

For runs with zero (or very low)
helicity and zero rotation, no clear 
evidence of $1/f$ noise is found, as the 
magnetic moments become essentially uncorrelated 
after about ten nonlinear times. 
The waiting time distributions occupy a much narrower
span of times, as reversals become numerous.
The distribution of waiting times does not match the observational 
data very well in this case, and 
appears to form a broad peak around a few nominal times. 
 
When runs are carried out with 
either rotation or helicity 
evidence for $1/f$ noise is found in all cases. 
In addition, the waiting times
begin to resemble the waiting times 
computed from the CK95 geomagnetic dataset.
When there is a more distinct spectral signature of 
$1/f$ power, one finds a better 
correspondence of the simulation waiting times
and the CK95 waiting times.
This is the case for numerical runs in which 
rotation is present, indicating that this effect is the most
important one to be considered to understand the reversals. 

In summary, the results from this series of runs show
that for reversals to have distributions of waiting times compatible
with known observational results it is necessary to have
rotation present.
Non-zero magnetic helicity is also an asset, but of less significance.
For these cases, 
the magnetic moment dipole has long time fluctuations, and a frequency
analysis shows a $1/f$-noise type spectrum. 
As reported in previous studies,
in these cases the turbulence readily generates
variability at very long dynamical time scales.

It is of course 
not possible to draw definite conclusions about the 
terrestrial dynamo from an oversimplified model
as the present one.
Even in the context of the model that we employ, 
the reported simulations 
have not been run with parameters as extreme
as those found in nature. For example, with eddy speed 
$1\times 10^{-8}$ km/s, $R\approx 3000$ km, and one rotation per day, 
a more realistic rotation parameter would be $\Omega t_0 \sim 10^5$.
But in that case runs extending to 5000 times $t_0$
would correspond to $\sim 10^9$ rotations. 
This would be a discouragingly stiff
numerical problem using our accurate (but computationally expensive) 
Galerkin code. Even then, longer runs, to perhaps 
$10^6 t_0$ would be required to compute reversals that might 
occur at $10^7$ years. 
In this perspective, the significance of the 
present results are largely due to the 
self-similar character of both the 
powerlaw waiting times and the apparently
underlying $1/f$ noise signal.
Conceptually, 
the present results 
demonstrate that 
the key physical ingredients 
present in a simple model of nonlinear 
magnetohydrodynamics, 
with rotation,
are able to 
account for statistics of reversals
roughly comparable to those observed 
for the terrestrial dynamo.
The requisite 
long timescales appear to 
originate 
in the $1/f$ noise generated by the 
model. This $1/f$ 
noise generation has been argued previously 
to be a generic feature of nonlinear systems operating 
in a regime on which nonlocality of interactions in scale 
is a prominent feature \cite{DmitrukMatthaeus07}.
As such, we suggest that
geomagnetic reversals
may in part share their physical origins
with a much broader class of nonlinear self-organizing 
fluid problems.

Research supported by grants PIP0825, 
UBACYT 20020110200359, PICT 2011-1529, 
2011-1626,
NSF AGS-1063439, SHINE AGS-1156094, CMG/1025183,
Solar Probe Plus Project through ISIS Theory team,
POR Calabria FSE 2007/2013 and
Marie Curie Project FP7 PIRSES-2010-269297 ``Turboplasmas''.

\end{document}